\documentclass{aastex631}

\usepackage{gensymb}

\def\Msun{\mbox{${\rm M}_{\odot}$}}
\def\Lsun{\mbox{${\rm L}_{\odot}$}}
\def\Rsun{\mbox{${\rm R}_{\odot}$}}

\received{November 12, 2021}

\shorttitle{Direct detections of {\rm sdO} companions orbiting classical {\rm Be} stars}
\shortauthors{Klement et al.}

\graphicspath{{./}{figures/}}

\begin{document}

\title{Interferometric detections of sdO companions orbiting three classical Be stars}

\author[0000-0002-4313-0169]{Robert Klement}
\affiliation{The CHARA Array of Georgia State University, Mount Wilson Observatory, Mount Wilson, CA 91023, USA}

\author[0000-0001-5415-9189]{Gail H. Schaefer}
\affiliation{The CHARA Array of Georgia State University, Mount Wilson Observatory, Mount Wilson, CA 91023, USA}

\author[0000-0001-8537-3583]{Douglas R. Gies}
\affiliation{Center for High Angular Resolution Astronomy, Department of Physics and Astronomy,\\ Georgia State University, P.O. Box 5060, Atlanta, GA 30302-5060, USA}

\author[0000-0003-4511-6800]{Luqian Wang}
\affiliation{Center for High Angular Resolution Astronomy, Department of Physics and Astronomy,\\ Georgia State University, P.O. Box 5060, Atlanta, GA 30302-5060, USA}
\affiliation{Yunnan Observatories, CAS, P.O. Box 110, Kunming 650011, Yunnan, China}
  
\author[0000-0003-1637-9679]{Dietrich Baade}
\affiliation{European Organisation for Astronomical Research in the Southern Hemisphere (ESO), \\ Karl-Schwarzschild-Str.\ 2, 85748 Garching bei M\"unchen, Germany}

\author[0000-0003-1013-5243]{Thomas Rivinius}
\affiliation{European Organisation for Astronomical Research in the Southern Hemisphere (ESO), Casilla 19001, Santiago 19, Chile}

\author[0000-0001-7853-4094]{Alexandre Gallenne}
\affiliation{Universidad de Concepci\'{o}n, Departamento de Astronom\'{i}a, Casilla 160-C, Concepci\'{o}n, Chile}
\affiliation{Unidad Mixta Internacional Franco-Chilena de Astronom\'{i}a (CNRS UMI 3386), Departamento de Astronom\'{i}a, Universidad de Chile, Camino El Observatorio 1515, Las Condes, Santiago, Chile}

\author[0000-0002-9369-574X]{Alex C. Carciofi}
\affiliation{Instituto de Astronomia, Geofísica e Ciências Atmosféricas, Universidade de São Paulo, Rua do Matão 1226}

\author[0000-0002-3380-3307]{John D. Monnier}
\affiliation{Department of Astronomy, University of Michigan, 1085 S. University Ave, Ann Arbor, MI 48109, USA}

\author[0000-0003-2125-0183]{Antoine Mérand}
\affiliation{European Organisation for Astronomical Research in the Southern Hemisphere (ESO), \\ Karl-Schwarzschild-Str.\ 2, 85748 Garching bei M\"unchen, Germany}

\author[0000-0002-2208-6541]{Narsireddy Anugu}
\affiliation{The CHARA Array of Georgia State University, Mount Wilson Observatory, Mount Wilson, CA 91023, USA}
\affiliation{Steward Observatory, Department of Astronomy, University of Arizona, 933 N. Cherry Ave, Tucson, AZ 85721, USA}

\author[0000-0001-6017-8773]{Stefan Kraus}
\affiliation{Astrophysics Group, School of Physics and Astronomy, University of Exeter, Stocker Road, Exeter, EX4 4QL, UK}

\author[0000-0001-9764-2357]{Claire L. Davies}
\affiliation{Astrophysics Group, School of Physics and Astronomy, University of Exeter, Stocker Road, Exeter, EX4 4QL, UK}

\author[0000-0001-9745-5834]{Cyprien Lanthermann}
\affiliation{The CHARA Array of Georgia State University, Mount Wilson Observatory, Mount Wilson, CA 91023, USA}

\author[0000-0002-3003-3183]{Tyler Gardner}
\affiliation{Department of Astronomy, University of Michigan, 1085 S. University Ave, Ann Arbor, MI 48109, USA}

\author[0000-0003-0392-1094]{Peter Wysocki}
\affiliation{Center for High Angular Resolution Astronomy, Department of Physics and Astronomy,\\ Georgia State University, P.O. Box 5060, Atlanta, GA 30302-5060, USA}

\author[0000-0002-1575-4310]{Jacob Ennis}
\affiliation{Department of Astronomy, University of Michigan, 1085 S. University Ave, Ann Arbor, MI 48109, USA}

\author[0000-0001-8837-7045]{Aaron Labdon}
\affiliation{Astrophysics Group, School of Physics and Astronomy, University of Exeter, Stocker Road, Exeter, EX4 4QL, UK}
\affiliation{European Organisation for Astronomical Research in the Southern Hemisphere (ESO), Casilla 19001, Santiago 19, Chile}

\author[0000-0001-5980-0246]{Benjamin R. Setterholm}
\affiliation{Department of Astronomy, University of Michigan, 1085 S. University Ave, Ann Arbor, MI 48109, USA}

\author[0000-0002-0493-4674]{Jean-Baptiste Le Bouquin}
\affiliation{Université Grenoble Alpes, CNRS, IPAG, 38000 Grenoble, France}

\begin{abstract}
Classical Be stars are possible products of close binary evolution, in which the mass donor becomes a hot, stripped O or B-type subdwarf (sdO/sdB), and the mass gainer spins up and grows a disk to become a Be star. While several Be+sdO binaries have been identified, dynamical masses and other fundamental parameters are available only for a single Be+sdO system, limiting the confrontation with binary evolution models. In this work, we present direct interferometric detections of the sdO companions of three Be stars 28~Cyg, V2119~Cyg, and 60~Cyg, all of which were previously found in UV spectra. For two of the three Be+sdO systems, we present first orbits and preliminary dynamical masses of the components, revealing that one of them could be the first identified progenitor of a Be/X-ray binary with a neutron star companion. These results provide new sets of fundamental parameters that are crucially needed to establish the evolutionary status and origin of Be stars.

\end{abstract}

\keywords{Be stars (142) -- O subdwarf stars (1138) --  Optical interferometry (1168) -- Orbit determination (1175) -- Multiple star evolution (2153)}

\section{Introduction} \label{sec:intro}

Classical B-emission line stars (hereafter referred to as Be stars) are rapidly rotating main sequence B-type stars surrounded by purely gaseous, ionized, self-ejected disks, and they represent about 15-20\% of all B-type stars in the local environment \citep[although this number can increase dramatically at low metallicities, see][for a review]{2013A&ARv..21...69R}. A rotation rate close to the critical value is a defining property of Be stars \citep[e.g.,][]{2016A&A...595A.132Z}, and it is one of the necessary conditions for the formation of circumstellar decretion disks, from which the eponymous line emission originates. The origin of the rapid rotation among Be stars therefore holds a key to understanding the Be star class as a whole. 

Binarity is a natural outcome of the star formation process. Components of close binary systems with periods shorter than a few years undergo periods of mass exchange that has a profound effect on the evolution \citep{1992ApJ...391..246P, 2014ApJ...782....7D} and the ultimate fate of these systems \citep[e.g.,][]{2020A&A...637A...6L}. Close binarity is observed to be common among massive stars \citep[see][for a review]{2013ARA&A..51..269D}. The fraction of close binaries among B-type stars in the Galaxy and the Large Magellanic Cloud appears to be at least 50\% based on interferometric \citep{2013MNRAS.436.1694R} and spectroscopic \citep{1990ApJS...74..551A, 2012Sci...337..444S, 2014ApJS..213...34K, 2015A&A...580A..93D, 2021MNRAS.tmp.1986V, 2021A&A...653A.144H} observations corrected for observational and selection biases. 

Based on simulations of binary-star populations, \citet{2013ApJ...764..166D} argued that rapidly rotating massive stars were spun up as a consequence of mass transfer or mergers, while \citet{2014ApJ...782....7D} came to the conclusion that $\sim30$\% of all main-sequence (MS) B-type stars are the products of binary interaction. Binary population synthesis studies specific to Be stars were also performed \citep[e.g.,][]{1991A&A...241..419P, 1997A&A...322..116V, 2014ApJ...796...37S, 2021A&A...653A.144H, 2021ApJ...908...67S}, suggesting that if not all, at least a significant fraction is probably made up of binary interaction products. Most Be stars formed by mass transfer in a binary should then have hot, subluminous companions of the subdwarf OB-type (sdOB) type (similar to isolated helium stars) that can further evolve into white dwarfs (WD) or neutron stars (NS) to form Be/X-ray binaries (BeXRB). Although black hole (BH) are also a possibility, the number of Be+BH systems is expected to be small \citep[e.g.][]{2015MNRAS.452.2773G}. 

On the other hand, even single stars could be created by close binary interaction, like the above mentioned case of fast-rotating merger products. The first identified case of a Be star being a merger product might be HD~93521, which is a rapidly rotating runaway late-O star with H$\alpha$ emission from a faint disk, whose location far from the galactic disk is incompatible with the evolutionary age if only single star evolution is considered. Thus, there is strong evidence that HD~93521 is the result of a merger of two lower-mass stars, which had formed a runaway binary system following a supernova-induced disruption of a triple system or a dynamical ejection (Gies et al., submitted). Supernova explosions of massive stripped companions in close binaries can also lead to systemic disruption, causing the Be stars to become single runaway stars \citep{2001ApJ...555..364B}. On the basis of the observed fraction of runaway stars in a large sample of Be stars ($\sim13\%$), \citet{2018MNRAS.477.5261B} conclude that all Be stars could be products of binary mass transfer.

Observationally, confirmed close binary systems comprise a non-negligible fraction of Be stars. Conspicuously, there is a lack of confirmed close MS companions to Be stars \citep{2020A&A...641A..42B}. As for stripped, evolved companions, the number of confirmed cases remains rather small. Other than the 200 or so known BeXRBs (which are however drawn from a large volume in the Local Group), there are only 15 spectroscopically confirmed cases of Be+sdO systems (no Be+sdB systems) and only a few more candidates (see below). Until recently, there were no confirmed cases of BeXRBs with WD (as opposed to NS or unconstrained) companions, despite the fact that detectable levels of low-luminosity X-ray emission had been expected \citep{1992A&A...265L..41M}. The elusive X-rays compatible with Be+WD systems, in which the WD accretes the outer Be disk material, were recently detected in the Magellanic Clouds for for a total of six viable candidates \citep[][and references therein]{2021MNRAS.508..781K}. A related case to Be+WD binaries is the Bn star (rapid rotator without a disk) Regulus, which is orbited by a faint pre-WD star \citep{2020ApJ...902...25G}. The one reported case of a Be star being orbited by a quiescent BH \citep{2014Natur.505..378C} relied on possibly overestimated measurements of radial velocities (RVs), and high-resolution spectra resemble the known cases of Be+sdO systems (Rivinius, unpublished). 

The low luminosity of sdO stars (compared to the Be primary) makes them difficult to detect spectroscopically in the visible (or at longer wavelengths), but their high $T_\mathrm{eff}$ makes their contribution more easily detectable in the far ultraviolet (FUV, see below). In the visible, a few sdO companions have been found by detecting antiphased motions of the \ion{He}{2}\,$\lambda 4686$ line (in absorption or emission) that is associated with the hot sdO star \citep[e.g.,][]{1981PASP...93..297P}. By heating up parts of the Be disk, the sdO can also give rise to single-peaked emission line components (e.g., in \ion{He}{1}\,$\lambda 6678$) that follow the orbital motion but with larger velocity amplitudes than those of the companion \citep[e.g.,][]{2000ASPC..214..581R}. Other spectroscopic features that may indicate the presence of a hot companion include for instance short-lived phases of narrow shell-line absorption, and short-term variations of the (typically double-peaked) strong emission lines like H$\alpha$ \citep{2005PAICz..93...21M}.

Importantly, direct detection of sdO companions was shown to be possible with near-IR long-baseline interferometry \citep[so far only for the case of $\varphi$~Per,][see below]{2015A&A...577A..51M}, a technique capable of achieving angular resolution equivalent to a (diffraction limited) aperture of hundreds of meters. One should note that recently, an interferometric multiplicity survey among bright ($m_V \leq 5.0$) Be stars (classical and non-classical) was performed using data from the NPOI optical interferometer, but in that case the dynamical range was insufficient for the detection of companions contributing less than $\sim4$\% in the visible (maximum $\Delta m$ of 3.5 at $\lambda=700$\,nm, \citealt{2021arXiv210906839H}). The study nevertheless provided the first direct detection of the helium/sdO star in the interacting binary $\upsilon$~Sgr (although the NPOI observations would be compatible also with an MS companion with a $\approx$ B5 spectral type). The helium star contributes $\sim4$\% of the total visible flux, despite being the more massive component in this peculiar system. 

Another possibility to detect companions of any kind is provided by their tidal influence on the Be star disk. With a close companion in orbit, the outer parts of the disk will be truncated, which translates into a lack of free-free emission at radio wavelengths \citep{1991A&A...244..120W, 2017A&A...601A..74K}. Analysis of the full sample of Be stars with published radio data led to the conclusion that all of them show the same characteristic turndown in the spectral energy distribution \citep[SED,][]{2019ApJ...885..147K}. The caveats of this method are that (1) the SED turndown could be caused by a hitherto unidentified physical effect that is separate from binarity, and that (2) the nature of the possible companion remains unconstrained. On the other hand, one can estimate the size of the orbit of the possible companion based on the measured size of the Be disk \citep{2015A&A...584A..85K}.

The first identified Be+sdO binary $\varphi$~Per ($P=127$\,d) was initially found from RV shifts of an emission component in the \ion{He}{2}\,$\lambda 4686$ line \citep{1981PASP...93..297P}. It was also the first one to be confirmed by cross-correlation techniques in FUV spectra, first using data from the International Ultraviolet Explorer short wavelength prime camera \citep[IUE/SWP,][]{1995ApJ...448..878T}, and later the Hubble Space Telescope \citep[HST,][]{1998ApJ...493..440G}. The sdO companion of $\varphi$~Per was also the first one - and until this study the only one - to be directly detected by long-baseline interferometry \citep{2015A&A...577A..51M}, and a combined spectroscopic and astrometric solution revealed that the sdO star has a mass of $1.2\pm0.2$\,\Msun. The sdO was found to contribute $\sim16$\% of the total flux in the FUV \citep{1998ApJ...493..440G}, but only $\sim1.5$\% of the total flux in the $H$-band (including the contribution from the Be star disk), and $\sim2.1$\% relative to the Be star photosphere alone \citep{2015A&A...577A..51M}. 

Several other Be+sdO systems were first detected (or suspected) from variable spectroscopic features before being confirmed in FUV spectroscopy. These are HR~2142 \citep{1983PASP...95..311P}, 59~Cyg \citep{2000ASPC..214..581R}, FY~CMa \citep{2004A&A...427..307R, 2005ASPC..337..309S}, and HD~55606 \citep{2018ApJ...865...76C}. The systems (also) found in archival IUE/SWP spectra include FY~CMa \citep{2008ApJ...686.1280P}, 59~Cyg \citep{2013ApJ...765....2P}, HR~2142 \citep{2016ApJ...828...47P}, and 60~Cyg \citep{2017ApJ...843...60W}. \citet{2018ApJ...853..156W} listed 12 additional IUE candidates, nine of which were subsequently confirmed in new FUV spectra from the HST Imaging Spectrograph (HST/STIS), while the three Be+sdO candidates that were not confirmed are cases where the sdO companions could be obscured by the Be disks during certain orbital phases \citep{2021AJ....161..248W}. A few additional Be+sdO candidates are suspected from the behavior of He lines in optical spectra but remain unconfirmed due to lack of FUV data: $o$~Pup \citep{2012MNRAS.424.2770V, 2012ASPC..464...75R, 2012A&A...545A.121K}, HD~161306 \citep{2014A&A...567A..57K}, 7~Vul \citep{2020A&A...639A..32H}, and HD 35165 (Rivinius, unpublished). It is worth noting that no sdO companion around a late-type Be star has been found \citep{2019IAUS..346..105R}, although the detection in the FUV spectra should be easier due to a lower FUV contribution from the Be star. This could be due to the fact that the search was optimized for hot sdO stars ($T_\mathrm{eff} \sim 45$\,kK), and failed to identify possible cooler sdB companions.

While there is still no unambiguously confirmed example of a Be star with a close MS companion \citep{1992IAUS..151..147B, 2000ASPC..214..668G, 2020A&A...641A..42B}, it should be mentioned that a significant number of Be stars have companions of unconfirmed nature. The SB1 systems 4\,Her, 88\,Her, $\epsilon$\,Cap, $\zeta$\,Tau, $\gamma$~Cas, and Pleione are a few examples of bright Be stars with companions that could be faint, late-type MS stars, thus challenging the binary scenario as the origin of all Be stars. It is worth noting that $\gamma$~Cas itself is a subject of intense study regarding its hard, variable X-ray emission with a luminosity of less that 1/30th of the values typical in BeXRBs but several times higher than in early B-type stars. These properties are shared by at least several more early-type Be stars, forming a subclass of $\gamma$~Cas stars \citep{2016AdSpR..58..782S, 2018A&A...619A.148N}. The origin of the X-ray emission and the connection to a possible prevalence of close binarity among $\gamma$~Cas stars remains unclear, as scenarios involving a WD, NS, sdO companion, as well as no companion at all, have been suggested \citep[see, e.g.,][and references therein]{2020A&A...633A..40L}. As for Be stars with likely MS companions, the two with the shortest orbital periods appear to be $\delta$~Cen \citep[period of $\sim$5.2\,yr, companion spectral type between B4\,V and A0\,III,][]{2008A&A...488L..67M, 2012A&A...538A.110M} and the wider inner pair in the 2+2 quadruple system $o$~And \citep[period of $\sim$5.7\,yr, companion spectral type estimated to be B6\,III,][]{2010ARep...54.1134Z}. The orbital periods in both of these are at least an order of magnitude larger that those of the known Be+sdO binaries. Another Be star with a possible MS companion is the highly eccentric $\delta$\,Sco with an orbital period of $\sim11$ years, which was suggested to be a runaway triple system \citep{2013ApJ...766..119M}. 

In this paper, we report on the first results from a survey of binarity among Be stars performed at the Center for High Angular Resolution Astronomy (CHARA) Array, which offers sub-milliarcsec angular resolution and is capable of detecting faint companions contributing as little as 0.2\% of the total flux in the near-IR \citep{2005ApJ...628..453T, 2020SPIE11446E..05S}. The targets are introduced in Sect.~\ref{sec:targets} and the observations are described in Sect.~\ref{sec:observations}. The new binary detections and preliminary orbits are presented in Sects.~\ref{sec:detections} and \ref{sec:preliminary_orbits} before concluding with Sect.~\ref{sec:conclusions}.

\section{Targets}
\label{sec:targets}

Basic information on the observed Be stars is shown in Table~\ref{tab:targets}, where the magnitudes $m_V$ and $m_H$ and the spectral types were adopted from the Simbad database\footnote{\url{http://simbad.u-strasbg.fr/simbad/}} \citep{2000A&AS..143....9W}. The estimates for angular limb-darkened (LD) diameters of the Be star photospheres were taken from \citet{2021AJ....161..248W} for 28~Cyg and V2119~Cyg and determined in the same way for 60~Cyg. The distances were taken from Gaia Early Data Release 3\footnote{\url{https://vizier.u-strasbg.fr/viz-bin/VizieR-3?-source=I/350}} (EDR3). Lastly, Table~\ref{tab:targets} lists the Be star physical radii $R_\mathrm{Be}$ computed from the LD diameter and the distance. 

\begin{deluxetable*}{llcccCCC}
\tablecaption{Sample Be stars\label{tab:targets}}
\tablewidth{0pt}
\tablehead{
\colhead{HD number} & \colhead{Name} & \colhead{$m_V$} & \colhead{$m_H$} & \colhead{Spectral type} & \colhead{LD diameter} & \colhead{Distance} & \colhead{$R_\mathrm{Be}$\tablenotemark{a}} \\
\nocolhead{HD number} & \nocolhead{Target} & \colhead{[mag]} & \colhead{[mag]} & \nocolhead{Spectral type} & \colhead{[mas]} & \colhead{[pc]} & \colhead{[\Rsun]}
}
\startdata
HD 191610 & 28~Cyg   & 4.9 & 5.2 &  B2.5\,Ve   &   0.214\pm0.010   &   256.9^{+7.7}_{-7.2} & 5.9\pm0.3 \\
HD 194335 & V2119~Cyg   & 5.9 & 6.2 &  B2\,IIIe    &   0.131\pm0.008   &   368.7^{+7.0}_{-6.7} & 5.2\pm0.3  \\
HD 200310 & 60~Cyg   & 5.4 & 6.0 &   B1\,Ve   & 0.122\pm0.004    &   375.6^{+18.4}_{-16.8} &  5.0\pm0.3  \\
\enddata
\tablenotetext{a}{Calculated from the angular diameter and the distance.}
\end{deluxetable*}

\subsection{28 Cyg}\label{sec:targets_28Cyg}

28~Cyg (HR\,7708, HD\,191610) currently possesses a developed if not particularly dense disk, as evidenced by amateur spectra available from the BeSS database\footnote{\url{http://basebe.obspm.fr}} \citep[a total of 351 H$\alpha$ spectra covering the period from 1995 to present,][]{BeSS}. At first sight, the double-peaked profile indicates that the circumstellar disk is seen under intermediate inclination. The equivalent width (EW) of the H$\alpha$ line has mostly stayed between $0$ and $-10$\,\AA, although there was a period of weakened emission (EW $\sim0$\,\AA) reflecting only a very weak disk between $\sim$2007 and $\sim$2012. \citet{1982ApJS...50...55S} reported long-term variability for H$\beta$ and H$\gamma$. While H$\beta$ was showing a symmetric, double-peaked emission in 1971 and 1979, it was in pure absorption in 1964. Interestingly, H$\gamma$ showed a `very weak flanking emission around a central absorption core' (in 1971) and a `sharp absorption core' (in 1979), suggesting a transient shell spectrum (i.e., disk seen close to edge-on inclination, cf. the case of 60~Cyg below).

28~Cyg has been extensively studied in the context of its short-term spectroscopic and photometric variations caused by non-radial pulsations (NRP), which are believed to be connected to the mass ejection mechanism that supplies material and angular momentum to the circumstellar disk \citep{2003A&A...411..229R, 2018A&A...610A..70B}. Variations attributable to a binary nature have not been reported until the orbiting sdO companion was found and its RVs measured by cross-correlation analysis of 46 archival IUE/SWP spectra \citep{2018ApJ...853..156W}. The sdO signature was detected (and its RVs measured) only in 25 out of the 46 exposures, and the resulting poor phase coverage prevented an estimation of the orbital period or any other orbital parameters. Although the companion was not confirmed by FUV spectroscopy from HST/STIS (three exposures) by \citet{2021AJ....161..248W}, the authors still consider it a likely Be+sdO system and suggest that an orbital phase-dependent obscuration of the companion by the Be star disk might be the reason why a detection was possible only at certain epochs. 

\subsection{V2119 Cyg}\label{sec:targets_V2119Cyg}
Judging from the 39 available BeSS spectra covering H$\alpha$ (1997--2021), V2119~Cyg (HR\,7807, HD\,194335) appears to have similarly evolving disk properties as \object{28\,Cyg}. While H$\alpha$ had EW around $-7$ to $-10$\,\AA\ until $\sim$2014, V2119~Cyg is currently found in a much weaker disk state with EW around $-2$\,\AA. Like in 28~Cyg, double-peaked H$\alpha$ suggests an intermediate inclination. According to \citet{1982ApJS...50...55S}, strong emission in H$\alpha$ and weak emission in H$\beta$ were present from 1953 to 1970. The emission in H$\beta$ apparently disappeared in 1975, but returned in 1979--80.

V2119~Cyg was placed among Be+sdO binary candidates based on cross-correlation analysis of four IUE/SWP spectra \citep{2018ApJ...853..156W}, which led also to RV measurements of the sdO. The sdO companion has recently been confirmed in three HST/STIS spectra \citep{2021AJ....161..248W}, making it possible to derive the following sdO parameters: $T_\mathrm{eff} = 43500$\,K, $v \sin{i} < 15$\,km\,s$^{-1}$, and flux contribution of $4.7\pm0.7$\% in the FUV \citep[][with three additional RVs measured for both the sdO and the Be star]{2021AJ....161..248W}. Using the observed SED, the authors also derived estimates for the radius and luminosity of the sdO star: $R = 0.52\pm0.07$\,\Rsun, and $\log{L} \mathrm{(\Lsun)} = 2.94^{+0.15}_{-0.23}$. Using the measured RVs (a total of seven measurements for the sdO star), \citet{2021AJ....161..248W} also presented a preliminary spectroscopic orbit with a period $P=60.286\pm0.010$\,d and a velocity semi-amplitude $K_2=75.5\pm2.4$\,km\,s$^{-1}$.  

\subsection{60 Cyg}\label{sec:targets_60Cyg}
60~Cyg (HR\,8053, HD\,200310) shows similar disk properties and long-term variability to both 28~Cyg and V2119~Cyg. The 74 available BeSS spectra covering H$\alpha$ (1995--present) show a diskless phase in the early 2000s, after which the emission gradually grew to EW of around $-10$\,\AA\ in 2013. After decreasing slightly to -7\,\AA\ in the few years after 2013, the EW grew to the present day value of around -12\,\AA. \citet{1982ApJS...50...55S} reported occasional `sharp dark cores' in Balmer lines from 1954 to 1976, decreasing H$\alpha$ emission between 1964 and 1976, and a double-peaked emission in H$\beta$ in 1979-80. As with 28~Cyg, this indicates a transient shell-like spectrum and the possibility of a close to edge-on inclination if the circumstellar disk.

Based on spectroscopic analysis, \citet{2000A&A...356..913K} suggested that 60~Cyg is a binary with a 146.6\,d period. They also derived a preliminary SB1 orbital solution based on RV variations of H$\alpha$ emission-line wings, with the resulting Be star $K_1$-velocity semiamplitude of $10.8\pm0.1$\,km\,s$^{-1}$, and a mass function $f(m)$ of 0.0191\,\Msun. 

60~Cyg was confirmed to be a Be+sdO system from cross-correlation analysis of 23 IUE/SWP spectra \citep{2017ApJ...843...60W}. From the detected sdO signature in the FUV spectra, the authors estimated a mass ratio $M_\mathrm{sdO}/M_\mathrm{Be} = 0.15\pm0.02$. Using the orbital solution and the Be star parameters derived by \citet{2000A&A...356..913K}, they also derived the flux contribution of the sdO star to be $3.39\pm0.15$\% in the FUV, and its mass and radius to be around $M=1.7$\,\Msun, and $R=0.48$\,\Rsun, respectively. Like \object{28\,Cyg}, \object{60\,Cyg} was also studied in the context of short period light and line-profile variations connected to NRP properties \citep{2000A&A...356..913K}.

\section{Observations}
\label{sec:observations}

The interferometric data of the three Be+sdO systems were collected at the CHARA Array \citep{2005ApJ...628..453T, 2020SPIE11446E..05S} as a part of an ongoing survey of binarity among Be stars that started in late 2020. The results presented here represent the first set of new binary detections and preliminary orbits from this survey. While we were able to derive preliminary orbits for two out of the three targets, we continue to monitor all three targets to arrive at fully constrained orbits that will lead to (more) precise estimates of dynamical masses. 
The CHARA Array consists of six 1-meter telescopes in a Y-shaped configuration \citep{2005ApJ...628..453T}. Two telescopes reside on each arm, and they are labeled according to the direction of the corresponding arm: two East telescopes E1 and E2, and correspondingly S1, S2, and W1, W2, for South and West, respectively. The longest baseline of CHARA - $B_\mathrm{max} \sim$330\,m - translates to an angular resolution of $\lambda / 2B_\mathrm{max} \sim 0.5$\,milliarcsec (mas) in the case of the $H$-band ($\lambda = 1.6$\,$\mu$m).

The Michigan InfraRed Combiner-eXeter (MIRC-X) installed at the CHARA Array is a highly-sensitive six-telescope interferometric image-plane beam combiner operating in the near-infrared $J$ and $H$ bands \citep{2020AJ....160..158A} with spectral resolution $R\sim50$ or $R\sim190$. Combination of all six telescopes results in a total of 15 visibility-squared (VIS2), 20 closure phases (CP), and 20 triple amplitudes (T3AMP). MIRC-X has demonstrated $H$-band sensitivity down to a correlated magnitude of 8.2 and VIS2 and CP precision of 1\% and 1\degree, respectively. It has also demonstrated the ability to detect companions with contrast ratios of up to 500:1 \citep{2020AJ....160..158A, 2015A&A...579A..68G}. A search in the Simbad database reveals that at the $H$-band magnitude limit of 8.2, MIRC-X can practically observe a sample of up to 383 Be stars.

The interferometric observing sequence consists of alternating between science targets and calibrators. The calibrators are stars with measured or well-estimated angular diameters, that are used to monitor the transfer function (i.e., the effect of atmospheric and instrumental bias on the interferometric observables) in between science observations, which is then used to calibrate raw science observables. Ideally, the calibrators are as close as possible to the science targets in terms of position on the sky, magnitude, and spectral type, while having a much smaller angular diameter. In practice, it is usually not possible to meet all of these conditions, so the observer has to compromise between the different factors.

The log of MIRC-X observations for our three science targets and the corresponding calibrators is shown in Table~\ref{tab:mircx_log}. All three targets were observed on at least three separate occasions between 2021 May and 2021 September. We used the $H$-band ($\lambda = 1.6$\,$\mu$m) and a low spectral resolution of $R\sim50$. The calibrators were selected using the SearchCal software \citep{2006A&A...456..789B, 2014ASPC..485..223B} provided by the Jean-Marie Mariotti Center (JMMC\footnote{\url{https://www.jmmc.fr}}), and their basic characteristics including the uniform disk (UD) diameter are listed in Table~\ref{tab:cals}. We used the official MIRC-X pipeline\footnote{\url{https://gitlab.chara.gsu.edu/lebouquj/mircx_pipeline.git}} (v.~1.3.5) to reduce the data \citep{2020AJ....160..158A}.

The interferometric field of view (FoV) is typically limited by the effect of bandwidth smearing, i.e., blurring of fringes with increasing spectral width of the individual wavelength channels. The bandwidth-smearing FoV is given by $\lambda/ (B\Delta \lambda)$, where $\lambda$ is the observed wavelength, $B$ is the baseline length, and $\Delta \lambda$ is the spectral width of an individual wavelength channel \citep{2015A&A...579A..68G}. For CHARA in the $H$-band with $\lambda/\Delta\lambda \sim 50$, bandwidth-smearing limits the FoV to maximum offsets from the phase center of $\sim50$\,mas. It is possible to correct analytically for the bandwidth smearing to a certain extent, but this was not needed for the present work.

\begin{deluxetable*}{lclCCcc}
\tablecaption{Log of MIRC-X observations\label{tab:mircx_log}}
\tablewidth{0pt}
\tablehead{
\colhead{Target} & \colhead{Average MJD} & \colhead{UT date} & \colhead{VIS2\tablenotemark{a}} & \colhead{CP\tablenotemark{a}} & \colhead{Configuration} & \colhead{Calibrators} \\
\nocolhead{UT Date} & \nocolhead{UT Time} & \nocolhead{Target} & \nocolhead{[min]} & \nocolhead{[min]} &  \nocolhead{Notes}
}
\startdata
28 Cyg & 59364.455 & 2021 May 30    & 570 & 561  & E1-W2-W1-S1-E2    & HD\,188256 \\
       & 59398.350 & 2021 Jul 3     & 432 & 608  & E1-W2-W1-S2-S1-E2 & HD\,188365 \\
       & 59399.337 & 2021 Jul 4     & 960 & 1280 & E1-W2-W1-S2-S1-E2 & HD\,188365, HD\,196881 \\
V2119 Cyg & 59397.424 & 2021 Jul 2  & 320 & 320  & E1-W2-W1-S2-E2    & HD\,188256, HD196881 \\
          & 59399.437 & 2021 Jul 4  & 960 & 1280 & E1-W2-W1-S2-S1-E2 & HD\,188256, HD\,196215 \\
          & 59439.344 & 2021 Aug 13 & 816 & 792  & E1-W2-W1-S2-S1-E2 & HD\,195647 \\
          & 59479.250 & 2021 Sep 22 & 440 & 528  & E1-W2-W1-S2-S1-E2 & HD\,196215, HD\,188256 \\
60 Cyg    & 59364.496 & 2021 May 30 & 343 & 249  & E1-W2-W1-S1-E2    & HD\,196881 \\
          & 59397.469 & 2021 Jul 2  & 640 & 640  & E1-W2-W1-S2-E2    & HD\,196881, HD204876\\
          & 59399.430 & 2021 Jul 4  & 1560 & 2080 & E1-W2-W1-S2-S1-E2 & HD\,196881, HD\,188256\\
          & 59439.396 & 2021 Aug 13 & 600 & 800  & E1-W2-W1-S2-S1-E2 & HD\,204876 \\     
          & 59479.215 & 2021 Sep 22 & 960 & 1256 & E1-W2-W1-S2-S1-E2 & HD\,188256, HD\,204876 \\
\enddata
\tablenotetext{a}{Total number of individual measurements.}
\end{deluxetable*}

\begin{deluxetable*}{lCCc}
\tablecaption{Calibrators\label{tab:cals}}
\tablewidth{0pt}
\tablehead{
\colhead{Target} & \colhead{UD diam. ($H$-band)\tablenotemark{a}} & \colhead{$m_H$} & \colhead{Spectral type} 
}
\startdata
HD 188256 & 0.322\pm0.007 & 5.88 & G5      \\
HD 188365 & 0.327\pm0.008 & 5.78 & G5      \\
HD 196215 & 0.415\pm0.009 & 5.45 & K0       \\
HD 195647 & 0.471\pm0.011 & 5.17 & K0\,III \\
HD 196881 & 0.306\pm0.007 & 5.98 & G5  \\
HD 204876 & 0.356\pm0.008 & 5.73 & G5     \\
\enddata
\tablenotetext{a}{Taken from the JMMC Stellar Diameters Catalogue v2 (\url{https://vizier.u-strasbg.fr/viz-bin/VizieR-3?-source=II/346}}
\end{deluxetable*}

\section{Binary detections}
\label{sec:detections}

The main interferometric observable VIS2 is a measure of the contrast of interference fringes formed from a combination of two separate beams. VIS2 measurements are sensitive to the geometrical size of the observed light source (along a projected baseline), and a resolved (i.e., larger than the given angular resolution) source will result in a decrease of the fringe contrast (visibility). The CP (phase of the bispectrum), on the other hand, results from a combination of three beams in a closed triangle, and is sensitive to departures from point symmetry in the interferometric FoV. In our case of searching for very faint companions that are offset from the phase center, CP is the more powerful observable, but VIS2 can provide important constraints on the geometrical extents and the flux ratio of the two components. CP is also more robust to unstable atmospheric conditions compared to VIS2 as it is invariant to atmospheric phase perturbations. An additional and not yet widely used observable is T3AMP (amplitude of the bispectrum), also resulting from a combination of three telescopes in a closed triangle, although this observable can be strongly affected by the atmosphere, similarly to VIS2 \citep[see, e.g.,][for an introduction to interferometric observables]{2011psi..book.....G, 2015poi..book.....B}.

A first look at the calibrated VIS2 reveals that the three targets are mostly unresolved (VIS2 $>0.8$) even on the longest baselines, corresponding to UD diameters of $\lesssim0.25$\,mas (cf. Table~\ref{tab:targets} with photospheric LD estimates). This prevents us from determining reliable angular extents of the Be stars, the angular orientation and tilt of their disks, or the relative flux contribution of the stellar photospheres and the circumstellar disks. Inspection of the calibrated CPs reveals that the signal remains within around $\pm5$\degree, ruling out strong asymmetries or the presence of a low contrast companion. As will be described below, the small CP signals unambiguously point towards the presence of very high contrast companions at separations of a few mas. 


We used the dedicated open-source code \textit{Companion Analysis and Non-Detection in Interferometric Data}\footnote{\url{https://github.com/amerand/CANDID}, \url{https://github.com/agallenne/GUIcandid}} \citep[CANDID,][]{2015A&A...579A..68G} to perform a search for binary companions around the three observed Be stars. As opposed to traditional grid search resulting in a $\chi^2$ map, CANDID uses a more rigorous approach of computing a 2D grid of fits, i.e., a multi-parameter fit is performed for each grid point which represents the initial guess for the position of the companion. With the primary component at the phase center, the parameters are the companion position ($\Delta\alpha$, $\Delta\delta$), the observed flux ratio of the companion and the primary $f_\mathrm{O} = f({\rm sdO}) / f({\rm Be})$, where the term $f({\rm Be})$ contains both the flux from the stellar photosphere and the surrounding disk (if present), and the UD diameter of the primary. In the fitting process, it is required that multiple starting points reach the same minimum, thus ensuring that the parameter space is fully explored and that the minima are not local. Based on the traveling distance (to convergence) of the fits in the parameter space, CANDID determines a posteriori the optimal grid step, certifying that a global minimum can be found by repeating the fitting process with the correct step size. CANDID also allows one to estimate the detection confidence level as a multiple of the standard deviation $\sigma$ (assuming that the data uncertainties are uncorrelated) with a maximum level of $\sigma=8.0$. Bootstrap sampling is used to derive the final best-fit parameter values and their uncertainties and to derive the companion position in the form of an error ellipse. CANDID also enables robust determination of a 3$\sigma$ detection limit on the flux ratio of undetected companions from a given dataset, which is done on the basis of `injecting' fake companions at different positions and with different flux ratios into the data. 

We searched for companions within the bandwidth-smearing FoV, i.e., up to a maximum separation of $\sim50$\,mas between the components. In a first step, we performed binary fits for the companion position and the flux ratio using CP only, and with the UD diameter of the primary Be star fixed at the estimated photospheric angular diameter listed in Table~\ref{tab:targets}. This solution resulted in unambiguous and highly significant ($>8.0\sigma$) detections of a high contrast companion at separations $<10$\,mas in all but one data set, namely 28~Cyg on MJD $=59398.350$ (with 3.8$\sigma$). However, despite having larger uncertainty, the less significant detection agrees well with the measurement from the night that followed (MJD $=59399.337$), so that the orbital point should still be valid. 

In subsequent steps, we included the VIS2 and T3AMP observables in the binary fits, while introducing the UD diameter of the Be star as a free parameter in the fit. In a few cases (see below), apparent small-scale miscalibration issues in VIS2 and T3AMP (probably caused by variable atmospheric conditions) affected the binary detections. First, the 60~Cyg observation on MJD~$=59439.395$ resulted in unusable VIS2 and T3AMP, but the CP signal remained unaffected. Since the binary detection resulting from CP only is consistent with the other detections for this object, we still consider this to be a reliable orbital point. Second, although including VIS2 confirms the above mentioned weaker detection (despite slightly lowering its significance) for 28~Cyg (MJD~$=59398.350$), including both VIS2 and in this case a rather noisy T3AMP results in a non-detection. Third, for another two measurements (MJD~$=59364.455$ and MJD~$=59479.215$), the detection confidence decreased slightly (to 7.2$\sigma$ and 5.0$\sigma$) after the inclusion of VIS2 (and remained similar after including also T3AMP), but the locations of the $\chi^2$ minima remained unambiguous and the flux ratios are consistent with the other measurements. 
 
Overall, all but the two above mentioned measurements (MJD~$=59439.396$ and MJD~$=59398.360$) resulted in consistent companion locations to within 3$\sigma$ when using CP only, CP+VIS2, and CP+VIS2+T3AMP. As the final values, we adopt the solution from a simultaneous fit to CP+VIS2, as it results in the most consistent flux ratio across the different epochs (standard deviation of $<0.1\%$). The one exception is the above mentioned date MJD~=~59439.396, where we use the result from fitting only the CP. The numbers are listed in Table~\ref{tab:detections}, where $\rho$ is the angular separation, PA is the position angle measured from North to East, and $\Delta$RA and $\Delta$DEC are the companion offsets. In the remaining columns, the uncertainties are given as 1$\sigma$ error ellipses with semi-major and semi-minor axes $\sigma$-$a$ and $\sigma$-$b$ and the PA of the semi-major axis $\sigma$-PA (measured from North to East). The uncertainty includes instrumental 0.5\% calibration error of the absolute wavelength scale \citep{2020AJ....160..158A}. An example plot with the final residuals is shown in Fig.~\ref{fig:residuals7}. The weighted averages of the observed flux ratios $f_\mathrm{O}=f\mathrm{(sdO)}/f\mathrm{(Be)}$ ($H$-band) are listed in Table~\ref{tab:fluxratio}. Here, the term $f\mathrm{(Be)}$ contains a possible contribution from the circumstellar disk of the Be star, and the $f_\mathrm{O}$ values thus possibly underestimate the flux ratio of the component photospheres only. 


Next, we turn to confronting the measured $H$-band flux ratios with previous results from FUV spectroscopy. The available photospheric flux ratios $f_\mathrm{O}$ (FUV) were measured through the amplitude of the cross-correlation function (CCF) of the combined and model subdwarf spectra \citep{2017ApJ...843...60W, 2021AJ....161..248W}. Since the FUV studies also led to preliminary estimates of the effective temperatures of the sdO companions, we can use the FUV flux ratio and model spectra of the components to determine the photospheric flux ratio as a function of wavelength and, in particular, find the calculated photospheric flux ratio $f_\mathrm{C}$ in the $H$-band. The model spectra used were from the TLUSTY OSTAR2002 grid \citep{2003ApJS..146..417L}, and from the TLUSTY BSTAR2006 grid \citep{2007ApJS..169...83L}, for the sdO and Be components, respectively. 

Compared to V2119~Cyg and 60~Cyg, the case of 28~Cyg is more problematic, as the spectral signature of the sdO component was absent in recent HST spectra \citep{2021AJ....161..248W}, even though it was previously detected in 25 of 46 archival spectra from IUE \citep{2018ApJ...853..156W}. This means that no estimate of the effective temperature, and only an upper limit on the flux ratio, could be derived \citep{2021AJ....161..248W}. Therefore, we assumed the temperature of the sdO component to be $T_{\rm eff} = 45$~kK (as found in other cases) and we estimated the flux ratio $f_\mathrm{O}$ in the FUV by comparing the average cross-correlation function peak height for 28~Cyg with those for other cases with a measured $f_\mathrm{O}$ (see the column with heading S/N in Table 2 of \citealt{2018ApJ...853..156W}).  

The measured and expected flux ratios are summarized in Table~\ref{tab:fluxratio}, which gives the flux ratio $f_\mathrm{O}$ (FUV) estimated from the FUV spectral analysis, the calculated flux ratio $f_\mathrm{C}$ in the $H$-band, and the weighted average of the observed flux ratios $f_\mathrm{O}$ in the $H$-band. Comparison of the model spectra with near-IR photometric measurements  \citep[2MASS,][]{2003yCat.2246....0C} revealed no significant contribution of the circumstellar disks to the total $H$-band flux at the time of the  observations (between the calendar years 1998 and 2000). Inspection of the available BeSS spectra shows that at that time, 28~Cyg had H$\alpha$ emission similar to the present day, V2119~Cyg had a stronger H$\alpha$ emission, and 60~Cyg, on the other hand, was found in an almost completely emission-less (and therefore diskless) state. Thus, direct comparison of $f_\mathrm{C}$ and $f_\mathrm{O}$ in the $H$-band should be possible at least for 28~Cyg and V2119~Cyg, while for 60~Cyg, the present-day disk contribution might lower the value of $f_\mathrm{O}$, thus underestimating the photospheric flux ratio. The agreement indeed turns out to be excellent for 28~Cyg and V2119~Cyg, but in the case of 60~Cyg, $f_\mathrm{C}$ (based on IUE spectra only) is $\sim1.6\%$ compared to $f_\mathrm{O}$ of $\sim1.2\%$, likely revealing the above mentioned contribution of the disk to the $H$-band flux. These numbers are similar to what was found for $\varphi$~Per (Sect.~\ref{sec:intro}), whose developed disk also contributes significantly to the $H$-band flux. Thus, our $H$-band results offer an important verification of the FUV detection methods and confirm the sdO nature of the companions.

Lastly, we derived upper limits on the flux contribution of additional undetected companions with the CANDID `injection' method. In this case, CANDID is used to remove analytically the previously detected companion from the observables before performing the injection, thus obtaining an unbiased detection level map for any potential additional companion that remains undetected. The average values of the limiting companion flux ratios in the $H$-band $f_\mathrm{lim}$ were calculated within a separation $\rho$ of 25 and 50\,mas from the primary component using the best-quality datasets for each target. Table~\ref{tab:detlims} shows the limiting flux ratio estimates for our three targets.

\begin{deluxetable}{lLCCCCCCCCCCC}
\tablecaption{Interferometric detections of sdO companions. \label{tab:detections}}
\tablewidth{0pt}
\tablehead{
\colhead{Star} & \colhead{MJD} & \colhead{$\rho$} & \colhead{PA}  & \colhead{$\Delta$RA} & \colhead{$\Delta$DEC} & \colhead{$\sigma$-$a$} & \colhead{$\sigma$-$b$} & \colhead{$\sigma$-PA} & 
\colhead{$\chi^{2}_\mathrm{red}$}\\ 
\nocolhead{Star} & \nocolhead{MJD} & \colhead{[mas]} & \colhead{[\degree]} & \colhead{[mas]} & \colhead{[mas]}& \colhead{[mas]} & \colhead{[mas]} & \colhead{[\degree]} & \nocolhead{n$\sigma$}
}
\startdata
\hline
28 Cyg & 59364.455\tablenotemark{a} & 4.140 & 199.485 & -1.381 & -3.903 & 0.034 & 0.023 & -87.78 & 1.25 \\ 
       & 59398.350\tablenotemark{b} & 6.321 & 164.550 &  1.684 & -6.093 & 0.057 & 0.049 & 58.82 & 1.33 \\ 
       & 59399.337 & 6.438 & 164.338 &  1.738 & -6.199 & 0.041 & 0.035 & 85.84 & 2.09 \\ 
V2119 Cyg & 59397.424 & 1.208 & 191.896 & -0.249 & -1.182 & 0.034 & 0.008 & 76.62 & 2.33 \\ 
          & 59399.437 & 1.174 & 210.386 & -0.594 & -1.013 & 0.023 & 0.015 & 36.36 & 1.86 \\ 
          & 59439.344 & 1.596 &  86.659 &  1.593 &  0.093 & 0.022 & 0.012 & -56.11 & 1.79 \\ 
          & 59479.250 & 1.826 & 300.107 & -1.578 &  0.915 & 0.034 & 0.015 & -87.13 & 1.41 \\ 
60 Cyg & 59364.496 & 2.377 & 187.663 & -0.317 & -2.356 & 0.045 & 0.017 & 80.49 & 1.11 \\ 
       & 59397.469 & 3.432 & 197.500 & -1.032 & -3.273 & 0.022 & 0.020 & -86.18 & 2.15 \\ 
       & 59399.430 & 3.333 & 198.109 & -1.036 & -3.168 & 0.025 & 0.018 & 34.92 & 2.81 \\ 
       & 59439.395\tablenotemark{c} & 0.858 & 359.332 & -0.010 &  0.858 & 0.026 & 0.015 & 77.46 & 0.92 \\ 
       & 59479.215\tablenotemark{a} & 1.219 &  29.908 &  0.608 &  1.057 & 0.031 & 0.020 & 71.43 & 2.27 \\ 
\enddata
\tablenotetext{a}{Slightly weaker detection confidence when including VIS2/T3AMP as opposed to using CP only (see text).}
\tablenotetext{b}{Marginal detection (see text).}
\tablenotetext{c}{No VIS2 or T3AMP available, the fit was done using CP only.}
\end{deluxetable}

\begin{deluxetable}{lccc}
\tablecaption{Measured (O) and calculated (C) sdO to Be flux ratios \label{tab:fluxratio}}
\tablewidth{0pt}
\tablehead{
\colhead{Star} & 
\colhead{$f_\mathrm{O}$ (FUV)} & 
\colhead{$f_\mathrm{C}$ ($H$-band)} & 
\colhead{$f_\mathrm{O}$ ($H$-band)\tablenotemark{a}} \\ 
\nocolhead{[\%]} & 
\colhead{[\%]} &
\colhead{[\%]} &
\colhead{[\%]}
}
\startdata
28 Cyg     &  $4.2 \pm 2.6$    &  $0.8 \pm 0.5$    &  $0.80 \pm 0.07$ \\
V2119 Cyg  &  $4.7 \pm 0.7$    &  $2.3 \pm 0.3$    &  $2.34 \pm 0.08$ \\
60 Cyg     &  $3.4 \pm 0.2$    &  $1.58 \pm 0.07$  &  $1.16 \pm 0.07$ \\
\enddata
\tablenotetext{a}{Due to possible disk contribution, these values might underestimate the photospheric flux ratios.}
\end{deluxetable}

\begin{deluxetable*}{lCCc}
\tablecaption{Detection limits on additional components\label{tab:detlims}}
\tablewidth{0pt}
\tablehead{
\colhead{Target} & \multicolumn{2}{c}{$f_\mathrm{lim}$ [\%]} \\
\nocolhead{Target} & \colhead{$\rho<25$\,mas} & \colhead{$\rho<50$\,mas}
}
\startdata
28~Cyg    & 0.58\pm0.06 & 0.63\pm0.06   \\
V2119~Cyg & 0.76\pm0.07 & 0.83\pm0.08   \\
60~Cyg    & 0.63\pm0.06 & 0.69\pm0.07   \\
\enddata
\end{deluxetable*}

\section{Preliminary orbits}
\label{sec:preliminary_orbits}

The changing companion positions on different dates are compatible with the expectation for binaries with orbital periods of one to a few months, and we clearly detect orbital motion in datasets separated by at least two nights (Table~\ref{tab:detections}). However, given the still very few orbital points, we can only make preliminary estimates of the orbits. The orbits described below were obtained with the Orbit Fitting Library\footnote{\url{http://www.chara.gsu.edu/analysis-software/orbfit-lib}} \citep{2006AJ....132.2618S, 2016AJ....152..213S}. 

The full set of parameters for a binary orbit includes the orbital period $P$, epoch of periastron $T$ (or $T_{\rm RV max}$ for circular orbits), eccentricity $e$, semimajor axis $a''$ in angular units and $a$ in physical units, orbital inclination $i$, longitude of the ascending node $\Omega$, longitude of periastron $\omega$, velocity semiamplitudes $K_1$ and $K_2$, and the systemic velocity $\gamma$. One can only get the apparent $a''$ from astrometric orbits, and the projected $a_1\sin{i}$ and $a_2\sin{i}$ (where $a=a_1+a_2$) only from spectroscopic orbits. Upon combination of astrometric orbits (to derive $i$) and double-lined spectroscopic orbits (to derive $a\sin{i}$), one can calculate the total mass using the 3rd Kepler law, and the individual masses using the ratios $ q = M_1/M_2 = K_2/K_1 = a_2/a_1$. When only a single-lined spectroscopic orbit (and therefore only either $a_1\sin{i}$ or $a_2\sin{i}$) is available, we need to know the distance to obtain $a$ from $a''$, after which we can calculate the total mass, as well as the individual masses from the knowledge of the ratio $a_1/a_2$ (see, e.g., \citealt{1973bmss.book.....B} for an introduction to binary orbits).

\subsection{28 Cyg}
\begin{figure}
\plottwo{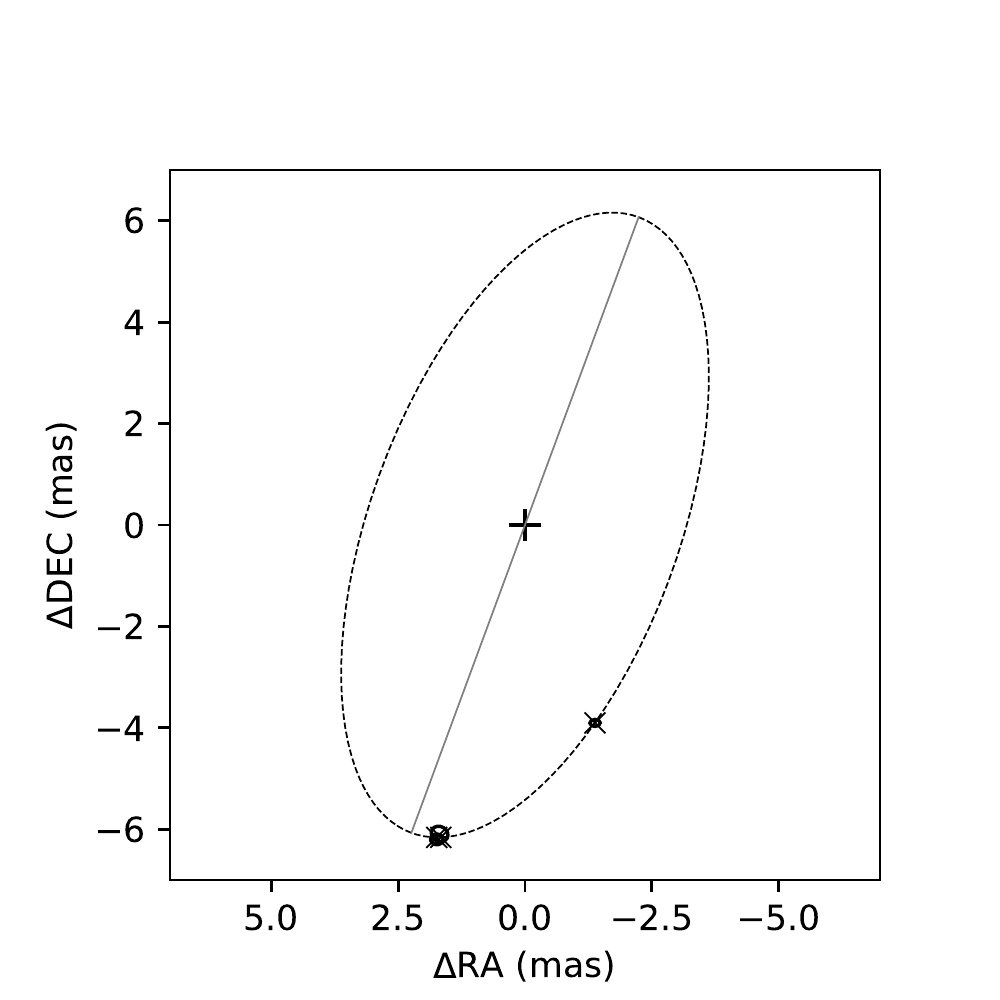}{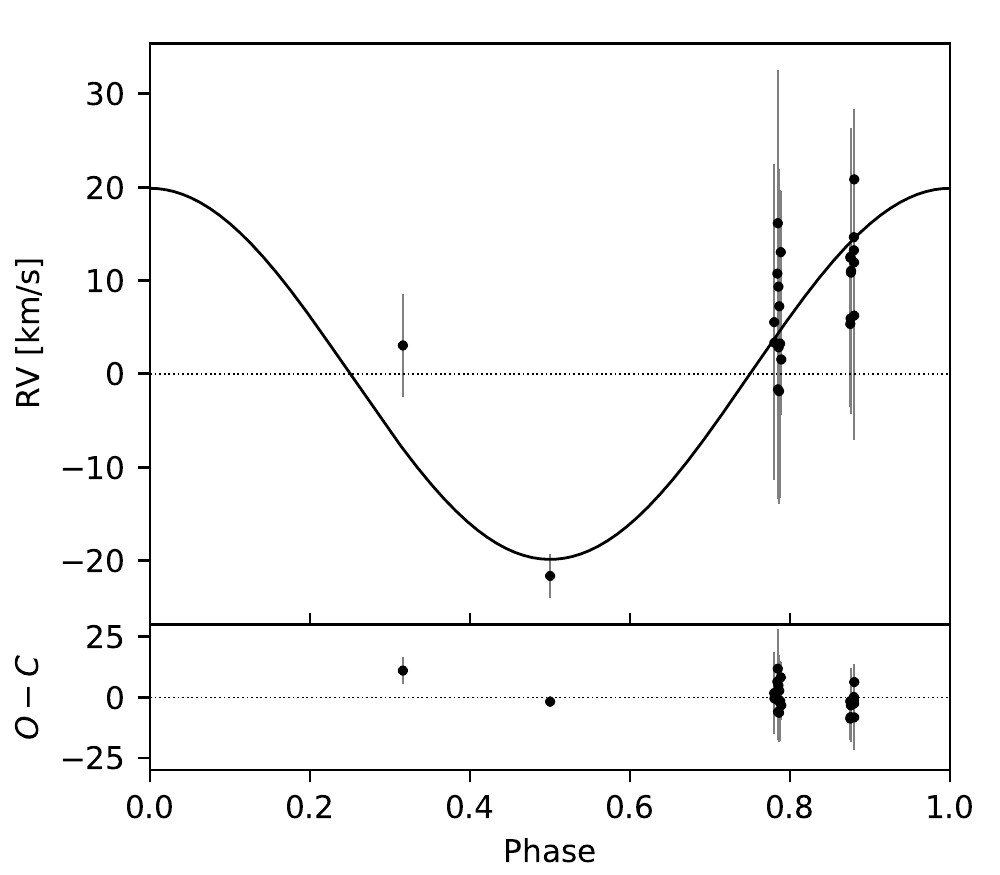}
\caption{\label{fig:28Cyg_orbit}Relative astrometric orbit and RV curve for the sdO companion of 28~Cyg. \textit{Left:} Possible (circular) orbit (dashed) of the sdO companion (with fixed $P=246$\,d) with the Be star at the center. The ellipses are the measured astrometric positions with 3$\sigma$ uncertainty and the corresponding points on the computed orbit are marked with x signs. The line of nodes is shown in grey. \textit{Right:} Possible RV curve of the secondary (solid line), corrected for the systemic velocity. RV measurements (black circles with grey error bars) are from \citet{2018ApJ...853..156W}. The lower panel shows the $(O-C)$ residuals.
}
\end{figure}

For 28~Cyg, no estimates for the orbit or orbital period are available in the literature. Although we have three interferometric detections, two of those were taken on subsequent nights and the results are compatible within the 3$\sigma$ error bars. The resulting coverage of the astrometric positions as well as that of the RVs measured by \citet{2018ApJ...853..156W} is thus insufficient for period estimation. Assuming that the measurement of angular separation of 6.44\,mas on MJD~=~59399.337 (Table~\ref{tab:detections}) is close to the orbital semimajor axis $a''$, and that the total mass of the system is 10\,\Msun\ \citep[based on the total mass of the $\varphi$~Per system with similar properties, ][]{2015A&A...577A..51M}, the orbital period $P$ resulting from the third Kepler law and Gaia EDR3 distance is $246$\,d. In Fig.~\ref{fig:28Cyg_orbit}, we plot a possible circular astrometric orbit and RV curve with $P$ fixed at this value. The resulting inclination for this orbit is $\sim118$\,\degree, indicating a retrograde (counter-clockwise) orbital motion. Fixing $P$ at lower values results in an increasingly edge-on inclination. Given that spectroscopic shell signatures were detected in the past (Sect.~\ref{sec:targets_28Cyg}), and assuming that the orbital and disk planes are aligned, a more inclined orbit is a possibility (cf. the case of 60~Cyg described below). The flux ratios are consistent across the three detections, not showing any strong variability that would be analogous to the FUV results, where the companion was detected only at certain epochs.

\subsection{V2119 Cyg}
For V2119~Cyg, \citet{2021AJ....161..248W} derived a preliminary circular SB1 orbit from the RVs of the sdO companion measured in FUV spectra. We used the available orbital parameters as the starting point in a simultaneous fit to the four astrometric positions and the six available RVs of the sdO companion. The result reveals a circular orbit seen at intermediate inclination with an excellent fit to both the astrometry and the RVs (Fig.~\ref{fig:V2119Cyg_orbit}). The resulting parameters are summarized in Table~\ref{tab:orb_pars_V2119Cyg}, where they are compared to those of \citet{2021AJ....161..248W}.

With the knowledge of the distance to the system, we can compute the physical size of the orbit as well as the masses of the components. The resulting mass of the sdO companion is $1.62\pm0.28$\,\Msun\, which is slightly above the Chandrasekhar limit ($\sim1.4$\,\Msun). Since isolated helium stars with masses above 1.6\,\Msun\ will ultimately explode as core-collapse supernovae \citep{2019ApJ...878...49W}, V2119~Cyg could be the first identified progenitor of a BeXRB with a NS companion \citep{2004ApJ...612.1044P}.

\begin{deluxetable*}{lCC}
\tablecaption{Parameters for V2119~Cyg\label{tab:orb_pars_V2119Cyg}}
\tablewidth{0pt}
\tablehead{
\nocolhead{} & \colhead{This study} & \colhead{Wang et al. (2021)} 
}
\startdata
$P$ [d] & 63.146\pm0.003   & 60.286\pm0.010   \\
$T_{\rm RV max}$ [RJD] & 58720.6\pm0.1  & 58720.7\pm0.6   \\
$e$ & 0\tablenotemark{a} & 0\tablenotemark{a}    \\
$\omega_\mathrm{Be}$ [\degree] & 0\tablenotemark{a}  & \nodata   \\
$K_\mathrm{Be}$ [km/s] & \nodata  & \nodata   \\
$K_\mathrm{sdO}$ [km/s] & 74.3\pm0.6  & 75.5\pm2.4   \\
$q$ & 0.20\pm0.03  & \nodata   \\
$i$ [\degree] & 49.4\pm0.5  & \nodata   \\
$\Omega$ [\degree] & 115.1\pm0.3  & \nodata   \\
$a''$ [mas] & 1.829\pm0.011  & \nodata   \\
$a$ [AU] & 0.676\pm0.015  & \nodata   \\
$M_\mathrm{Be}$ [\Msun] & 8.65\pm0.35  & \nodata   \\
$M_\mathrm{sdO}$ [\Msun] & 1.62\pm0.28  & \nodata   \\
$\gamma$ [km/s] & -18.4\pm0.5 & -18.9\pm1.2
\enddata
\tablenotetext{a}{Fixed for a circular orbit.}
\end{deluxetable*}

\begin{figure}
\plottwo{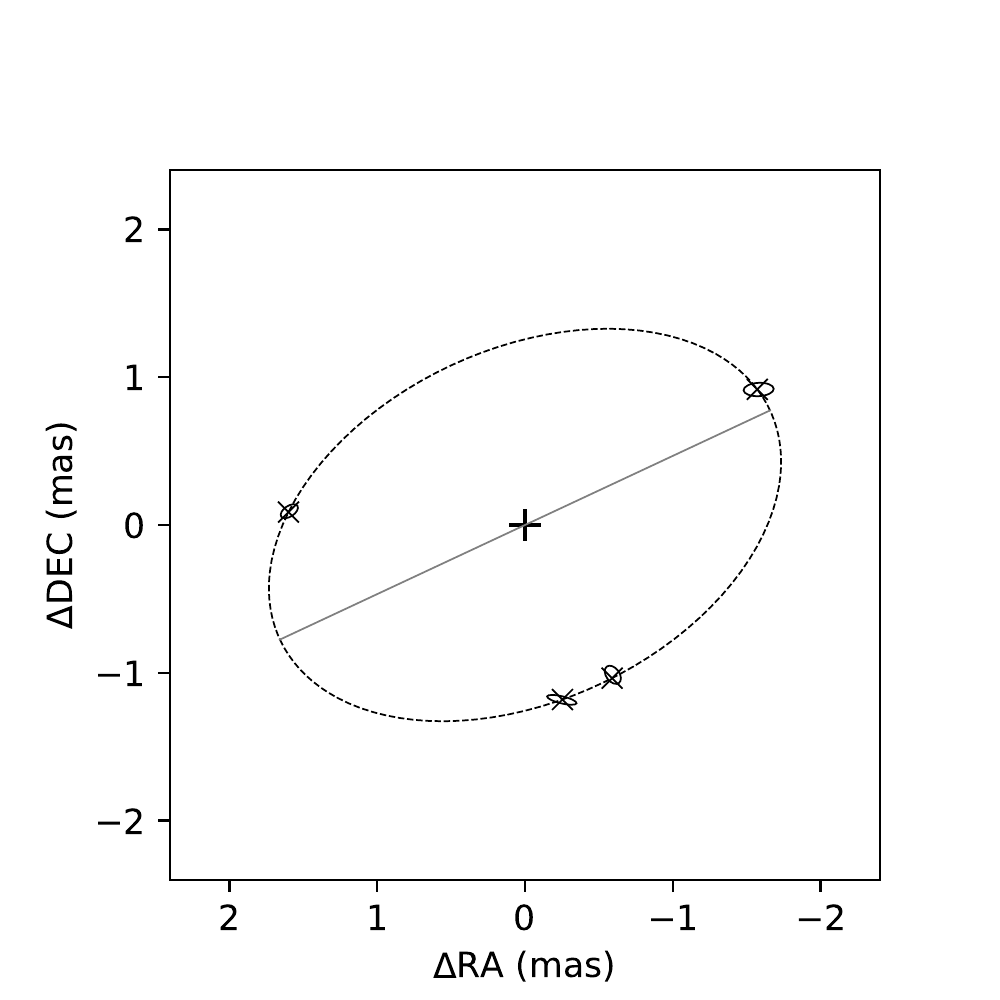}{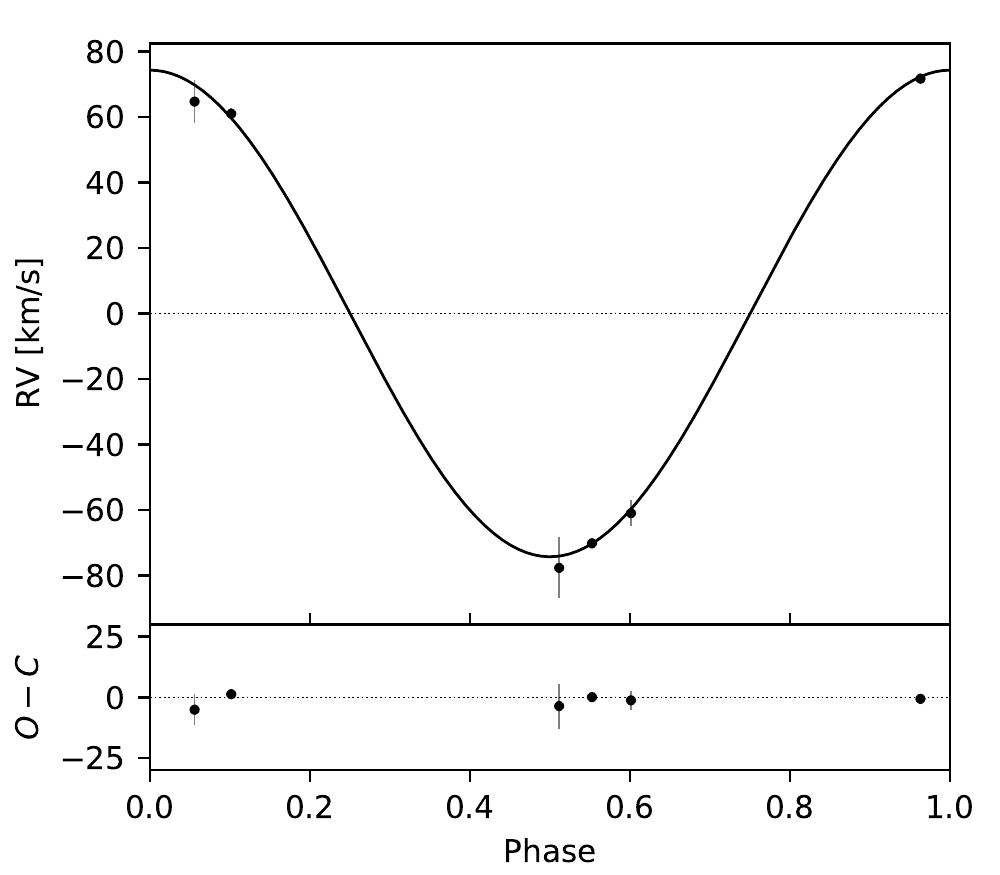}\label{fig:V2119Cyg_orbit}
\caption{\label{fig:V2119Cyg_orbit}Relative astrometric orbit and RV curve for the sdO companion of V2119~Cyg (symbols are the same as in Fig.~\ref{fig:28Cyg_orbit}). \textit{Left:} Preliminary (circular) orbit of the sdO companion resulting from a joint astrometric and RV fit, with the Be star at the center. \textit{Right:} RV curve of the secondary with RV measurements from \citet{2018ApJ...853..156W, 2021AJ....161..248W}.
}
\end{figure}

\subsection{60 Cyg}
For 60~Cyg, a preliminary single-lined spectroscopic orbit was given by \citet{2000A&A...356..913K}. We started with a fit to the astrometric positions while keeping the spectroscopic parameters fixed according to the available solution to obtain first guesses of the other orbital parameters. Since we were not able to find a satisfactory circular orbit, we allowed the eccentricity to vary, and arrived at a slightly eccentric orbit ($e\sim0.2$) seen at high inclination ($i>80$\degree). Subsequently, we added the 47 RVs of the Be primary (measured from H$\alpha$ wings by \citealt{2000A&A...356..913K}), for which we arbitrarily adopted errors of $\pm5$\,km/s. The scatter of the RV measured is large, showing the difficulty of measuring precise RVs of Be stars. The resulting orbit and RV curve are plotted in Fig.~\ref{fig:60Cyg_orbit}. As with V2119~Cyg, we can calculate the physical size of the orbit and preliminary dynamical masses of both components (with knowledge of the distance). The final parameters are listed in Table~\ref{tab:orb_pars_60Cyg}, where they are compared to results from previous studies, namely $P$, $T_\mathrm{RVmax}$, $K_\mathrm{Be}$, $\gamma$ and $M_\mathrm{Be}$ derived by \citet{2000A&A...356..913K}, and $q$ and $M_\mathrm{sdO}$ derived by \citet{2017ApJ...843...60W}, revealing a good agreement except for the component masses, which are lower according to our solution. We note that our preliminary Be star dynamical mass of $7.3\pm1.1$ also appears to be low for a B1-type star according to available spectroscopic calibrations based on dynamical masses \citep[e.g.][]{1988BAICz..39..329H}, although this might be simply due to a misclassification of the spectral type. 

Assuming that the disk orientation is aligned with the binary orbit, the high orbital inclination of $83$\degree\ seems to be at odds with the non-shell spectral appearance. Rather, the available spectra in the BeSS database suggest an inclination around 70-75\degree (cf. Fig.~1 in \citealt{1996A&AS..116..309H}). However, signatures of shell absorption have been observed in the past (Sect.~\ref{sec:targets_60Cyg}), which indicates that either (1) the disk is seen close to edge-on, but shell absorption is for some reason not generally apparent, or that (2) the orientation of the disk oscillates in time similarly to the cases of Pleione, $\gamma$~Cas, and 59~Cyg, although much less dramatically \citep[e.g.][]{2007PASJ...59L..35T}. Considering the first option, the shell absorption could be diluted by emission from an increased amount of disk material at zero RV. This would be similar to the case of the shell star $\zeta$~Tau, where a density wave propagates through its disk (manifesting itself via periodic asymmetries of the H$\alpha$ emission), and as the low density region passes in front of the star (at zero RV), the shell absorption is diluted by emission from the high density region behind the star (also at zero RV, \citealt{2009A&A...504..915C}). In Be stars without strong density waves in the disk, like 60~Cyg \citep{2000A&A...356..913K}, a similar effect would occur if the vertical height of the disk was enhanced, so that there is enough disk material next to the star (or rather below or above) that is not projected onto the stellar disk (cf. Fig.~5 in \citealt{1995A&A...295..423H}). 

Considering the non-zero eccentricity of the resulting orbit, there is at least one other known case of a slightly eccentric Be+sdO system, for which it was suggested that it might be caused by the dynamical influence of a third wide companion \citep{2013ApJ...765....2P}. Thus, we speculate that also 60~Cyg could have a wide companion, although it is not apparent at the separations and flux ratios probed by CHARA/MIRC-X data (cf.~Table~\ref{tab:detlims}).

\begin{deluxetable*}{lCC}
\tablecaption{Parameters for 60~Cyg\label{tab:orb_pars_60Cyg}}
\tablewidth{0pt}
\tablehead{
\nocolhead{} & \colhead{This study} & \colhead{Previous studies} 
}
\startdata
$P$ [d] & 147.68\pm0.03 & 146.6\pm0.6   \\
$T$ [RJD] & 59009.5\pm0.6 & 50016.9\pm1.9\tablenotemark{a}   \\
$e$ & 0.207\pm0.003 & 0\tablenotemark{b}    \\
$\omega_\mathrm{Be}$ [\degree] & 155.4\pm1.6 & \nodata   \\
$K_\mathrm{Be}$ [km/s] & 11.6\pm1.2 & 10.8\pm0.1   \\
$K_\mathrm{sdO}$ [km/s] & \nodata & \nodata   \\
$q$ & 0.164\pm0.037 & 0.15\pm0.02   \\
$i$ [\degree] & 83.4\pm0.3 & \nodata   \\
$\Omega$ [\degree] & 196.6\pm0.2 & \nodata   \\
$a''$ [mas] & 2.971\pm0.009  & \nodata   \\
$a$ [AU] & 1.12\pm0.06  & \nodata   \\
$M_\mathrm{Be}$ [\Msun] & 7.3\pm1.1  & 11.8^{+0.9}_{-1.0}   \\
$M_\mathrm{sdO}$ [\Msun] & 1.2\pm0.2  & 1.7   \\
$\gamma$ [km/s] & -13.0\pm0.6 & -13.4
\enddata
\tablenotetext{a}{The literature value is $T_{\rm RV max}$}
\tablenotetext{b}{Fixed for a circular orbit.}
\end{deluxetable*}

\begin{figure}
\plottwo{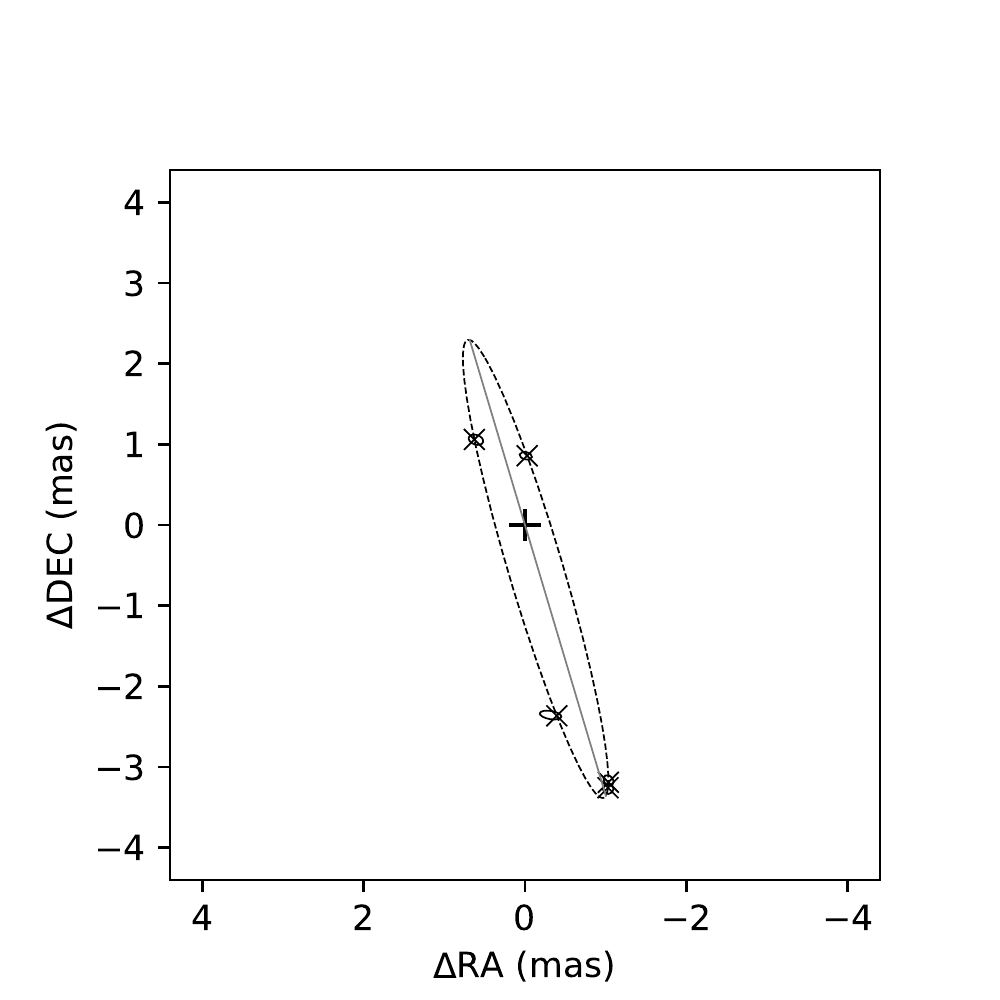}{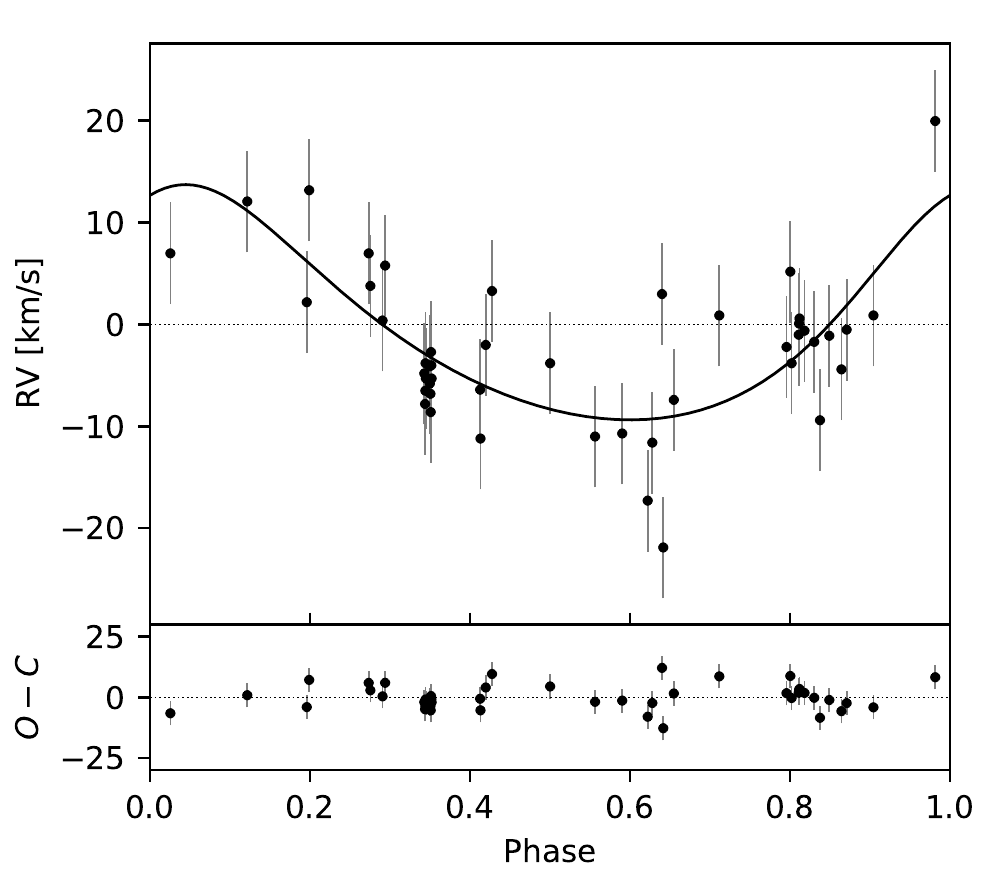}\label{fig:60Cyg_orbit}
\caption{\label{fig:60Cyg_orbit}Relative astrometric orbit and RV curve for the sdO companion of 60~Cyg (symbols are the same as in Fig.~\ref{fig:28Cyg_orbit}). \textit{Left:} Preliminary (slightly eccentric) orbit of the sdO companion resulting from a joint astrometric and RV fit, with the Be star at the center. \textit{Right:} RV curve of the primary with RV measurements from \citet{2000A&A...356..913K}.
}
\end{figure}

\section{Conclusions}
\label{sec:conclusions}

For the first time, sdO companions of the classical Be stars \object{28\,Cyg}, \object{60\,Cyg}, and \object{V2119\,Cyg} were directly detected with optical interferometry, confirming the Be+sdO nature in the case of 28~Cyg through an excellent agreement with the expected near-IR flux ratio (Table~\ref{tab:fluxratio}), and bringing the total of confirmed Be+sdO binaries to 16 (22 when including strong candidates). This represents only the second set of interferometric detections of (classical) Be+sdO binaries, the first one being the single case of $\varphi$~Per \citep{2015A&A...577A..51M}. These results demonstrate the usefulness of optical interferometry for binary detection and the derivation of astrometric orbits at angular resolution inaccessible to any single-telescope observing technique.

The positions of the detected companions are consistent with orbital periods ranging from one to a few months (Table~\ref{tab:detections}). Preliminary orbital solutions were derived from the new astrometry and available RV measurements for V2119~Cyg and 60~Cyg (Tables~\ref{tab:orb_pars_V2119Cyg} and \ref{tab:orb_pars_60Cyg}), while for 28~Cyg we were only able to suggest a possible orbit due to insufficient number of interferometric measurements. Preliminary dynamical masses for V2119~Cyg and 60~Cyg indicate that in the case of V2119~Cyg, the sdO companion might be massive enough ($1.62\pm0.28$) to explode in a core-collapse supernova, and therefore that V2119~Cyg itself could be the first identified progenitor of a BeXRB with a NS component. For 60~Cyg, the resulting orbit is slightly eccentric, making it possibly the second Be+sdO system whose orbit is not circular (after 59~Cyg). Ongoing mapping of the orbits and monitoring in high-resolution spectroscopy are expected to lead to refinement of the orbits and the dynamical masses in the near future, providing crucial and robust test cases for binary evolution models.

The measured flux ratios in the $H$-band were compared with those derived from previous FUV results and model spectra, revealing a good agreement (Table~\ref{tab:fluxratio}). 
The interferometric $H$-band results are thus an important verification of the FUV methods used to detect the companions, and to derive their flux ratios as well as fundamental parameters such as $T_\mathrm{eff}$.   


\begin{acknowledgments}
This work is based upon observations obtained with the Georgia State University Center for High Angular Resolution Astronomy Array at Mount Wilson Observatory. The CHARA Array is supported by the National Science Foundation under Grant No. AST-1636624, and AST-2034336. Institutional support has been provided from the GSU College of Arts and Sciences and the GSU Office of the Vice President for Research and Economic Development.
MIRC-X received funding from the European Research Council (ERC) under the European Union's Horizon 2020 research and innovation programme (Grant No.\ 639889) and N.A., C.L.D., and S.K.\ acknowledge funding from the same grant.
R.K. is grateful for a postdoctoral associateship funded by the Provost’s Office of Georgia State University. The research of R.K. is also supported by the National Science Foundation under Grant No. AST-1908026.
R.K. expresses sincere thanks to the CHARA science staff, namely Matthew D. Anderson, Theo ten Brummelaar, Christopher Farrington, Robert Ligon, Olli Majoinen, Nic Scott, Judit Sturmann, Lazslo Sturmann, Nils Turner, and Norman Vargas (only those not in the author list listed here), for making the observations used in this study possible.
A.L. received support from an STFC studentship (No.\ 630008203).
This research has made use of the SIMBAD database, operated at CDS, Strasbourg, France
A.G acknowledges support from the ALMA -ANID fund No. ASTRO20-0059.
A.C.C. acknowledges support from CNPq (grant 311446/2019-1) and FAPESP (grants 2018/04055-8 and 2019/13354-1). 
J.D.M. acknowledges NSF AST-2009489.
This research has made use of the Jean-Marie Mariotti Center \texttt{Aspro} and \texttt{SearchCal} services (Available at http://www.jmmc.fr/aspro and http://www.jmmc.fr/searchcal).
\end{acknowledgments}

\facilities{CHARA}

\bibliography{MIRC-X_Be+sdO_detections.bib}

\appendix

\section{An example plot showing the final residuals}

\restartappendixnumbering

\begin{figure}[h!]
   \centering
   \includegraphics[width=1.0\textwidth]{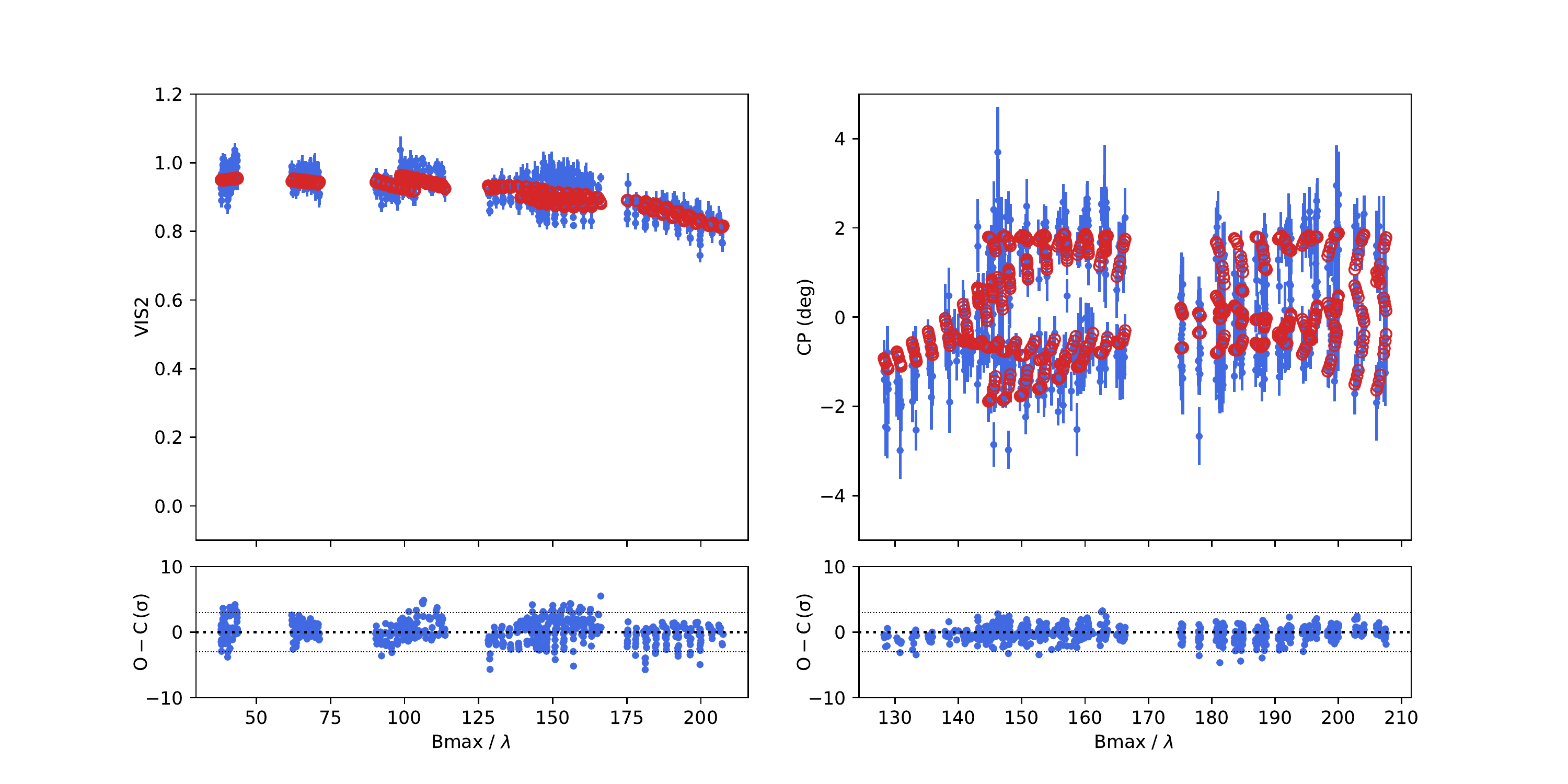}%
   \caption{\label{fig:residuals7}Example of a binary model fit (red) to the interferometric data (CP and VIS2, blue) and the resulting $O-C$ residuals (lower panels) for 60~Cyg observed on 2021 Jul 02 in 5-telescope mode. The thin grey dashed lines in the lower panels indicate the level of $\pm3\sigma$.
   }
\end{figure}

\end{document}